\documentclass[DIV=15,11pt,abstract]{scrartcl}

\usepackage[utf8]{inputenc}
\usepackage{amssymb, mathtools, bm, amsthm, thmtools, relsize}
\usepackage{float}
\usepackage{graphicx} 
\usepackage{adjustbox}

\usepackage[bottom]{footmisc}   

\usepackage[toc,page]{appendix}
\usepackage[noadjust]{cite} 
\usepackage{xcolor}
\definecolor{myBlue}{RGB}{1,1,141}

\usepackage{graphicx}
\usepackage{subcaption} 
\usepackage{authblk}

\usepackage[colorlinks=True]{hyperref}
\hypersetup{allcolors=myBlue}

\newcommand{\cH}{\mathcal{H}}

\renewcommand{\max}{\text{max}}

\newcommand{\ket}[1]{\left| #1 \right\rangle}
\newcommand{\bra}[1]{\left\langle #1 \right|}
\newcommand{\braket}[2]{\left\langle #1 \middle| #2 \right\rangle}

\title{Coordinate space representation for quantum simulation of scalar field theory}

\author[1]{Gaétan~Bardy}

\author[2]{Matthieu~Saubanère}

\author[1]{Adrian~Tanasa}

\affil[1]{\textit{Université de Bordeaux, LaBRI CNRS UMR 5800, Talence, France}}
\affil[2]{\textit{Université de Bordeaux, LOMA CNRS UMR 5789, Talence, France}}

\date{\today}

\begin{document}

\maketitle

\begin{abstract}
Quantum computing provides a promising framework for the simulation of quantum field theories, where the computational cost depends both on the quantum algorithm employed and on the representation of the Hamiltonian.
We investigate a formulation of the $\phi^4$ model based on the harmonic-oscillator basis in coordinate space.
We derive the lattice $\phi^4$ Hamiltonian in this representation and analyze the structure of the resulting one-body matrix and interaction tensor. We show that both exhibit an effective band-diagonal structure, allowing controlled truncations of the Hamiltonian while preserving the low-energy spectrum. We validate this formulation by comparing low-energy observables obtained from numerical diagonalization with those computed in the standard harmonic-oscillator momentum-space representation.
Finally, we estimate the resources required to encode the Hamiltonian on a quantum computer using both binary and unary boson-to-qubit mappings. 
By exploiting effective locality, the coordinate-space representation reduces the resources required for quantum simulation over a broad range of parameters.

\end{abstract}

\section{Introduction}

In a fundamental physics setting, quantum field theory  provides the natural framework for the description of relativistic many-body systems. It underlies the Standard Model of particle physics and also plays a central role in condensed matter physics, where it provides an effective description of critical phenomena and phase transitions.

One of the simplest interacting quantum field theoretical models is the $\phi^4$ model, which describes a single self-interacting scalar field. This celebrated model captures many essential features of interacting field theories and serves as a standard benchmark for analytical and numerical methods. Beyond its role as a toy model, the $\phi^4$ model has numerous physical applications, including the description of critical phenomena, models of cosmic inflation, and the Higgs sector of the Standard Model 
(see for example the book \cite{Kleinert2001CriticalPhi4}). 

Let us emphasize here that analytical treatments of interacting quantum field theories are generally limited to the perturbative regime, making non-perturbative methods, which are much harder, essential for studying strongly correlated systems.
Due to these difficulties, 
considerable effort has been devoted to the development of numerical approaches 
in order to analyse
quantum field theoretical models. Among these, lattice field theory provides a systematic regularization of the continuum theory and has become one of the standard frameworks for non-perturbative calculations. In particular, lattice Quantum Chromodynamics  has achieved remarkable success in the computation of hadronic observables, including hadron masses, and now constitutes the reference approach for obtaining predictions within the Standard Model (see for example \cite{L_QCD_1, L_QCD_2, L_QCD_3}).

Several alternative numerical methods have also been applied to the study of the $\phi^4$ model. This includes Monte Carlo techniques \cite{MonteCarlo1}, matrix product state (MPS) methods \cite{MPS1, MPS2} or the density matrix renormalization group (DMRG) \cite{DMRG2, DMRG1}. Another successful approach is the Hamiltonian truncation, see for example \cite{Lee_2001, Rychkov_2015, Bajnok_2016, Elias_Mir_2017,  Elias_Mir__2020, fitzpatrick2022snowmasswhitepaperhamiltonian}.
For example, in \cite{Rychkov_2015}, Rychkov and Vitale
 regularize the $\phi^4$ model by introducing a sharp energy cutoff $\Lambda$ in the Fock space rather than by discretizing the space itself. The Hamiltonian is then projected onto the finite-dimensional subspace spanned by all Fock states with energy below $\Lambda$. This allows the low-energy spectrum to be computed using exact diagonalization or iterative eigensolvers such as the Krylov subspace methods.
 
Let us emphasize that a necessary step before any such numerical analysis is the regularization of the infinite number of degrees of freedom of the theory. These regularization parameters (usually, a lattice spacing $a$ or an energy cutoff $\Lambda$) prevent the divergences (that appear when performing analytical computations) 
to appear in this setting. 

Despite their success, all of these 
classical numerical 
approaches eventually face an exponential growth of computational resources, either through the size of the truncated Hilbert space or, for tensor-network methods, through the growth of entanglement \cite{MPS_DMRG1}. This ultimately limits the range of parameters and system sizes that can be studied in practice.

These challenges faced by classical simulation methods have motivated the exploration of quantum computers as a tool to simulate quantum many-body systems and fundamental physical theories \cite{Feynman, Zalka1996st}. Quantum computers have entered the noisy intermediate-scale quantum (NISQ) era \cite{NISQ}, with a variety of hardware platforms enabling both digital quantum computation and analog quantum simulation (see, for example, \cite{archi1, archi2, archi3}). A wide range of quantum algorithms has been developed for high-energy physics applications, including Hamiltonian simulation, state preparation, and scattering calculations (see the reviews \cite{review1, review2, review3, review4, review5}).

Following the seminal work of Jordan, Lee, and Preskill \cite{Jordan_2012, jordan2019quantumcomputationscatteringscalar}, substantial effort has been devoted to the development of quantum algorithms for quantum field theory simulations. This includes advances in the discretization and encoding of quantum fields \cite{PhysRevA.99.052335, SinglePart, Macridin_2022, farrelly2020discretizingquantumfieldtheories}, as well as the development of algorithms for state preparation, Hamiltonian simulation, and ground-state estimation \cite{VarJLP, LimRessources, Pave, Liu_2020, ingoldby2024enhancingquantumfieldtheory, ingoldby2025realtimescatteringquantumcomputers}. Proof-of-concept implementations targeting existing quantum hardware, including IBM quantum processors, have also been proposed \cite{Wstates, SCVQE}.

Furthermore, note that, beyond their physical relevance, quantum field theories also provide valuable benchmarks for quantum computers and it has also been related to
computational complexity, with several quantum field theoretical simulation tasks proved to be BQP-complete \cite{Yeter_Aydeniz_2019, BQP}.

Let us emphasize that, independently of the quantum algorithm considered, the computational cost is strongly influenced by the structure of the Hamiltonian after encoding onto qubits. In particular, the number of Pauli strings and the Pauli $1$-norm provide useful measures of the resources required for its implementation \cite{costScale1norm}. For example, in variational quantum eigensolver (VQE) algorithms, the number of measurements required to estimate the energy depends on the number of Pauli terms, while in block-encoding approaches the query complexity typically scales linearly with the Pauli $1$-norm.
These structural properties, including sparsity and locality, depend on the choice of basis used to represent the Hamiltonian. Consequently, selecting an appropriate representation prior to applying a quantum algorithm can reduce the resources required for the simulation. Similar considerations have motivated the study of optimized basis choices in other quantum simulation problems, such as quantum chemistry \cite{basisSetforChemistry, TransfoFor1norm}.
In the context of quantum field theory, this motivates the investigation of alternative representations of the Hamiltonian where locality and sparsity can be exploited to reduce the qubit implementation cost.

Two types of basis are commonly used for 
quantum simulations of 
scalar field theories.
    The first such basis is the field-amplitude basis in coordinate space 
    in which the field amplitude is discretized independently at each lattice site.
    The second communly used basis is the harmonic-oscillator basis in momentum space,
    which corresponds to the basis obatined through the standard  canonical quantization of the scalar field. 
    We denote this basis by HO${}_p$. 

 In the field-amplitude basis, the free-theory vacuum is a multidimensional Gaussian state whose preparation on a quantum computer generally requires a non trivial quantum circuit, which can be obtained using the Kitaev-Webb algorithm \cite{KW}. Working in the harmonic-oscillator basis diagonalizes the free Hamiltonian, but the interaction term becomes highly non-local in momentum space, leading to long-range couplings in its qubit representation.

\medskip

In this paper, we investigate the encoding of the $\phi^4$ Hamiltonian in the harmonic-oscillator basis and coordinate-space representation, which we refer to as HO${}_x$.
We give numerical evidence that, although neither the free nor the interacting Hamiltonian is exactly local in this representation, both exhibit an effective band-diagonal structure in the presence of a mass gap.
This means that we can then efficiently implement a quantum algorithm in order to simulate the 
$\phi^4$ model, which constitute a highly interesting perspective for future work.

We further compare this formulation with the HO${}_p$ representation of the model in terms of the number of qubits required, the number of Pauli strings appearing in the Hamiltonian decomposition, and the Pauli $1$-norm. Moreover, we compare two standard boson-to-qubit mappings, namely the unary and binary encodings . 

The implementation cost of the Hamiltonian in the HO${}_p$ representation  has previously been analyzed. In ~\cite{Pave}, the authors estimate the resources required to implement both the Hamiltonian and its time-evolution operator. We use these results as 
a benchmark for comparison with the HO${}_x$ representation analyzed in this work. In addition, we compare our low-energy spectra with those obtained using the Hamiltonian truncation approach of Ref.~\cite{Rychkov_2015} in order to verify that both formulations converge to the same asymptotic results.

Our main results are summarized in Table~\ref{tab:sumup_intro}. We find that the HO${}_x$ representation admits effective bandwidth cutoffs for both the one-body and interaction terms, thus reducing the number of Pauli strings required to represent the Hamiltonian in both unary and binary encodings. For a fixed Hilbert-space size and couplings close to the critical regime ($\lambda_0 \simeq 3$), we further find that the Pauli $1$-norm of the coordinate-space Hamiltonian is smaller than that of its momentum-space counterpart.
\begin{table}[h]
\centering
\begin{adjustbox}{max width=\linewidth}
\begin{tabular}{|l|cc|cc|}
\hline
& \multicolumn{2}{c|}{\textbf{Binary encoding}} &
\multicolumn{2}{c|}{\textbf{Unary encoding}} \\
\cline{2-5}
& \textbf{$H_p$-binary} &
\textbf{$H_x$-binary} &
\textbf{$H_p$-unary} &
\textbf{$H_x$-unary} \\
\hline
$N_q$ &
$(2N_\mathrm{max}+1)\log_2(N_\phi)$ &
$N_s\log_2(N_\phi)$ &
$(2N_\mathrm{max}+1)N_\phi$ &
$N_sN_\phi$ \\
\hline
$N_\mathrm{Paulis}$ &
$\lesssim(2N_\mathrm{max}+1)^4N_\phi^8$ &
$\lesssim C_hN_sN_\phi^4 + C_UN_sN_\phi^8$ &
$\lesssim (2N_\mathrm{max}+1)^4N_\phi^4$ &
$\lesssim C_hN_sN_\phi^2 + C_UN_sN_\phi^4$ \\
\hline
\end{tabular}
\end{adjustbox}
\caption{
Summary of the resource requirements for encoding the $\phi^4$ Hamiltonian in the HO basis in momentum space ($H_p$) and position space ($H_x$), using binary and unary encodings. 
Here, $N_s$ denotes the number of lattice sites and $N_{\max}=N_s/2$ the maximum momentum mode. 
$N_\phi$ denotes the local Hilbert space dimension. 
The parameters $C_h$ and $C_U$ represent the effective bandwidths of the one-body and interaction terms, respectively, in the position-space representation. 
The position-space formulation reduces the scaling of the number of Pauli strings with the number of modes/sites by exploiting the effective locality of the Hamiltonian, independently of the bosonic encoding.
}
\label{tab:sumup_intro}
\end{table}

The paper is organized as follows. In Sec.~\ref{sec2}, we review the construction of the lattice $\phi^4$ Hamiltonian in the field-amplitude basis in coordinate space and in the HO${}_p$ representation. In Sec.~\ref{sec3}, we express the  $\phi^4$ Hamiltonian in the HO${}_x$ representation. In Sec.~\ref{sec4}, we analyze the structural properties of the resulting Hamiltonian and show that both the one-body and interaction terms can be accurately approximated by band-diagonal tensors. In the following section we compare its low-energy spectrum with that obtained in the momentum-space formulation and demonstrate that the two approaches converge to the same asymptotic results. In Sec.~\ref{sec5}, we estimate the quantum resources required to implement the Hamiltonian in both coordinate and momentum space, and compare unary and binary boson-to-qubit encodings. Finally, Sec.~\ref{sec6} gives our conclusion and lists various perspectives for future work.

\section{Review of the $\phi^4$ model}
\label{sec2}
In this section we review the scalar field theory in $(1+1)d$ on a circular lattice. We recall the Hamiltonian and the field digitization, in both amplitude and harmonic oscillator basis.

\subsection{The $\phi^4$ model in the amplitude basis in coordinate space}

The continuum Hamiltonian can be decomposed as
\begin{equation}
    H = H_0 + H_\lambda,
\end{equation}
where the free Hamiltonian is given by
\begin{equation}
    H_0
    =
    \int \mathrm{d}^d x
    \left(
    \frac{1}{2}\Pi^2
    + \frac{1}{2}(\nabla \phi)^2
    + \frac{m^2}{2}\phi^2
    \right),
\end{equation}
with $m$ the bare mass and $\Pi$ the canonical conjugate momentum to $\phi$. The interaction term reads
\begin{equation}
    H_\lambda
    =
    \frac{\lambda}{4!}
    \int \mathrm{d}^d x
    \phi^4,
\end{equation}
where $\lambda$ denotes the bare coupling constant. The field operator $\phi(\mathbf{x})$ and its conjugate momentum $\Pi(\mathbf{x})$ satisfy the canonical equal-time commutation relations. The quadratic part $H_0$ describes a free scalar field, while the quartic interaction $H_\lambda$ introduces self-interactions between field excitations. 
We discretize the theory on a one-dimensional periodic lattice consisting of $N_s$ sites with lattice spacing $a$, such that the total system size is $L=aN_s$. The resulting lattice Hamiltonian is
\begin{equation}
\label{ham}
H = a \sum_j
\left[
\frac{1}{2}\Pi_j^2
+
\frac{1}{2}(\nabla_a \phi_j)^2
+
\frac{m^2}{2}\phi_j^2
+
\frac{\lambda}{4!}\phi_j^4
\right].
\end{equation}
Here, $\phi_j$ denotes the field amplitude at site $j$, while $\Pi_j$ is the corresponding canonical conjugate momentum. The lattice introduces an ultraviolet cutoff $\Lambda_{\mathrm{UV}}\sim \pi/a$, while the finite system size provides an infrared cutoff of order $2\pi/L$.

The spatial derivative is approximated by a finite difference,
\begin{equation}
    (\nabla_a \phi_j)^2
    =
    \left(
    \frac{\phi_{j+1}-\phi_j}{a}
    \right)^2,
\end{equation}
with periodic boundary conditions imposed through $\phi_{N_s+1}\equiv \phi_1$.

The operators $\phi_j$ and $\Pi_j$ form a canonically conjugate pair at each lattice site. We denote by $\mathcal{F}$ the discrete Fourier transform acting on the local Hilbert space associated with a single lattice site. The conjugate momentum operator is then given by
\begin{equation}
    \Pi_j = \mathcal{F}^\dagger \phi_j \mathcal{F}.
\end{equation}
The field operators and their conjugate momenta satisfy the canonical commutation relations
\begin{equation}
\label{CCRs1}
[\phi_i,\Pi_j]=ia^{-1}\delta_{ij}, \qquad
[\phi_i,\phi_j]=[\Pi_i,\Pi_j]=0.
\end{equation}
These relations define a set of coupled quantum harmonic oscillators, with the quartic term introducing local self-interactions at each lattice site.
A quantum state in this representation is specified by a field configuration on the lattice,
\begin{equation}
    \ket{\varphi}
    =
    \ket{\phi_1,\ldots,\phi_{N_s}}.
\end{equation}

As already mentioned in the Introduction, the field basis serves as the starting point for many quantum algorithms for simulating scalar field theories (see again, for example \cite{jordan2019quantumcomputationscatteringscalar,VarJLP,LimRessources,SCVQE}). In practice, the local field degree of freedom must be regularized by introducing both a field cutoff $\phi_{\mathrm{max}}$ and a discretization step $\delta_\phi$. The allowed field values at each site are
\begin{equation}
    \phi_j^{(b)}
    =
    -\phi_{\mathrm{max}}
    +
    b\,\delta_\phi,
    \qquad
    b=1,\ldots,N_\phi,
\end{equation}
where $N_\phi$ denotes the number of discretized field values per lattice site. The resulting Hilbert space has dimension
\begin{equation}
    \dim(\mathcal H)
    =
    N_\phi^{N_s}.
\end{equation}

The ground-state preparation in this representation is generally demanding. Even in the free theory, the vacuum state is a multidimensional Gaussian with correlations extending across the lattice. Preparing such a state therefore requires generating entanglement between lattice sites, which typically dominates the state-preparation cost and one needs, for example to use the Kitaev-Webb algorithm.

Note that this already allow us to mention an advantage of the choise of representation we propose in this paper, because, as we will see in the sequel, the ground state of the free theory in the HO${}_x$ representation
is much easier to prepare (see Section \ref{sec3} below).

\subsection{Harmonic-oscillator basis in momentum space}

An alternative discretization is obtained by expanding the field operators in momentum modes and introducing the corresponding harmonic-oscillator ladder operators. 
Following the canonical quantization procedure, the field and conjugate momentum operators are expanded as
\begin{equation}
\begin{split}
    \phi_j
    &=
    \sum_n
    \frac{1}{\sqrt{2L\omega_n}}
    \left(
    a_n e^{i\frac{2\pi n}{N_s}j}
    +
    a_n^\dagger e^{-i\frac{2\pi n}{N_s}j}
    \right),
    \\
    \Pi_j
    &=
    -i
    \sum_n
    \sqrt{\frac{\omega_n}{2L}}
    \left(
    a_n e^{i\frac{2\pi n}{N_s}j}
    -
    a_n^\dagger e^{-i\frac{2\pi n}{N_s}j}
    \right).
\end{split}
\end{equation}
Here, $n \in \left[-N_{\max},\,N_{\max}\right]$, with $N_{\max}=\frac{N_s-1}{2}$ labels the momentum modes of the reciprocal lattice and $\omega_n$ denotes the lattice dispersion relation,
\begin{equation}
    \omega_n
    =
    \sqrt{
    m^2
    +
    \frac{4}{a^2}
    \sin^2\!\left(
    \frac{\pi a n}{L}
    \right)
    }.
    \label{omega}
\end{equation}
The operators $a_n$ and $a_n^\dagger$ are interpreted as annihilation and creation operators for the momentum mode $n$. Substituting the mode expansion into the canonical commutation relations (\ref{CCRs1}) yields
\begin{equation}
     [a_n,a^\dagger_{n'}]
     =
     \,\delta_{n,n'},
     \qquad
     [a_n,a_{n'}]
     =
     [a_n^\dagger,a_{n'}^\dagger]
     =
     0.
     \label{CCRs2}
\end{equation}
The Hilbert space of the free theory is described in the Fock basis. A quantum state is specified by the occupation number of each momentum mode,
\begin{equation}
    \ket{\mathbf{r}}
    =
    \ket{
    r_{-N_{\mathrm{max}}},
    \ldots,
    r_{N_{\mathrm{max}}}
    }.
\end{equation}
The integers $r_n \geq 0$ denote the number of excitations in mode $n$. These states are generated by repeated application of the creation operators on the vacuum state,
\begin{equation}
    \ket{\mathbf{r}}
    =
    \prod_n
    \frac{(a_n^\dagger)^{r_n}}
    {\sqrt{r_n!}}
    \ket{0}.
\end{equation}
The ladder operators act independently on each mode and can be written as
\begin{equation}
    a_n
    =
    \sum_{r=0}^{\infty}
    \sqrt{r+1}\,
    \ket{r}_n \bra{r+1}_n,
\end{equation}
where $\ket{r}_n$ denotes the $r$-particle state of mode $n$. Similarly,
\begin{equation}
    a_n^\dagger
    =
    \sum_{r=0}^{\infty}
    \sqrt{r+1}\,
    \ket{r+1}_n \bra{r}_n.
\end{equation}

In practice, in order to simulate the model on a (classical or quantum) computer, the infinite-dimensional Hilbert space associated with each harmonic oscillator must be truncated by imposing a maximum occupation number $(N_\phi-1)$. The resulting finite-dimensional Fock space has dimension
\begin{equation}
    \mathrm{dim}(\cH) = N_\phi^{2N_{\mathrm{max}}+1}.
\end{equation}
Substituting the mode expansion into the lattice Hamiltonian and normal ordering the resulting expression yields a free Hamiltonian that is diagonal in the HO${}_p$ representation, 
\begin{equation}
    H_0
    =
    \sum_n
    \omega_n a_n^\dagger a_n,
\end{equation}
where $\omega_n$ is given by eq.~\eqref{omega}. 
The interacting contribution takes the form
\begin{equation}
\begin{split}
    H_\lambda
    &=
    \frac{\lambda}{4!}
    \sum_{n_1+n_2+n_3+n_4=0}
    \frac{1}{\prod_i \sqrt{2L\omega_i}}
    \Big[
    a_{n_1}a_{n_2}a_{n_3}a_{n_4}
    +
    4a^\dagger_{-n_1}a_{n_2}a_{n_3}a_{n_4}
    \\
    &\qquad
    +
    6a^\dagger_{-n_1}a^\dagger_{-n_2}a_{n_3}a_{n_4}
    +
    4a^\dagger_{-n_1}a^\dagger_{-n_2}a^\dagger_{-n_3}a_{n_4}
    +
    a^\dagger_{-n_1}a^\dagger_{-n_2}a^\dagger_{-n_3}a^\dagger_{-n_4}
    \Big].
    \label{H_lambda_pspace}
\end{split}
\end{equation}
The constraint
\begin{equation}
    n_1+n_2+n_3+n_4=0
\end{equation}
expresses momentum conservation and follows from the translational invariance of the theory. 
Let us mention that, following for example \cite{Rychkov_2015},
throughout this paper, we restrict the Hilbert space to the zero-momentum sector.
The normal-ordering procedure also generates additional counterterms. Furthermore, finite-volume effects induce corrections to the Hamiltonian parameters. These contributions are exponentially suppressed in both the volume and the mass scale \cite{Rychkov_2015} and may therefore be neglected in the regime $ mL \gg 1$.

Let us end this section by noticing that, in the HO${}_{p}$ representation, 
the free vacuum is the Fock vacuum annihilated by the operators $a_n$, and the free Hamiltonian is diagonal.
Consequently, this representation provides a particularly efficient description of the weak-coupling regime. On the other hand, the interaction term couples many momentum modes simultaneously, resulting in a highly non-local Hamiltonian whose numerical implementation becomes increasingly costly as the coupling strength is increased.
Note that this allow us to mention here another advantage of the choise of representation we propose in this paper, because, as we will see in the sequel, the interacting Hamiltonian is the HO${}_x$ representation
is 
band diagonal
(see the following section).

\section{The $\phi^4$ model in the harmonic oscillator basis in coordinate space}
\label{sec3}

In this section we define the Hamiltonian of the $\phi^4$ model in the HO${}_x$ representation. 

As already mentioned above, while the momentum-space harmonic-oscillator basis diagonalizes the free Hamiltonian, it is also possible to construct an harmonic oscillator basis directly in coordinate space. One can 
consider the action of the field operator on the vacuum. Using the mode expansion introduced above,
\begin{equation}
    \phi_j \ket{0}
    =
    \sum_n
    \frac{1}{\sqrt{2L\omega_n}}
    a_n^\dagger
    e^{-i\frac{2\pi n}{N_s}j}
    \ket{0}.
\end{equation}
The state $\phi_j\ket{0}$ may be interpreted as a localized excitation centered around the lattice site $j$, see  \cite{Jordan_2012, jordan2019quantumcomputationscatteringscalar}. This observation naturally leads to the definition of localized creation and annihilation operators,
\begin{equation}
    a_j
    =
    \sum_n
    \frac{1}{\sqrt{2L\omega_n}}
    a_n
    e^{i\frac{2\pi n}{N_s}j},
    \qquad
    a_j^\dagger
    =
    \sum_n
    \frac{1}{\sqrt{2L\omega_n}}
    a_n^\dagger
    e^{-i\frac{2\pi n}{N_s}j}.
\end{equation}
These operators reduce to the usual Fourier transform in the non-relativistic limit, where the mode dependence of $\omega_n$ becomes negligible.

Their commutation relations are given by
\begin{equation}
\begin{split}
    [a_j,a_k^\dagger]
    &=
    \frac{1}{L}
    \sum_{n,m}
    \frac{
    e^{-i\frac{2\pi n}{N_s}(mk-nj)}
    }
    {2\sqrt{\omega_n\omega_m}}
    [a_n,a_m^\dagger]
    \\
    &=
    \frac{1}{L}
    \sum_n
    \frac{1}{2\omega_n}
    e^{-i\frac{2\pi n}{N_s}(k-j)}
    \\
    &=
    G_0(k-j),
\end{split}
\end{equation}
where $G_0$ is the free two-point correlation function on the lattice.
Unlike the momentum-space operators, the localized operators do not satisfy canonical bosonic commutation relations. Although $G_0(k-j)$ decays exponentially with the distance $|k-j|$ in a gapped theory, it is not strictly proportional to $\delta_{jk}$. Consequently, the many-body states generated by repeated applications of the operators $a_j^\dagger$ are not exactly orthogonal.


The non-orthogonality of the localized operators $a_j^\dagger$ originates from the mode-dependent factor $1/\sqrt{\omega_n}$ appearing in their definition. 
In order to recover canonical bosonic commutation relations, 
we introduce a different set of localized operators defined through the standard lattice Fourier transform,
\begin{equation}
    b_j
    =
    \frac{1}{N_s}\sum_n
    e^{i\frac{2\pi n}{N_s}j}
    a_n,
    \qquad
    b_j^\dagger
    =
     \frac{1}{N_s} \sum_n
    e^{-i\frac{2\pi n}{N_s}j}
    a_n^\dagger.
    \label{bj}
\end{equation}
These operators preserve the Fock vacuum,
\begin{equation}
    b_j\ket{0}=0,
    \qquad
    \forall j,
\end{equation}
and satisfy canonical commutation relations. Indeed,
\begin{equation}
\begin{split}
    [b_j,b_k^\dagger]
    &=\frac{1}{N_s^2}
    \sum_{n,m}
    e^{i\frac{2\pi}{N_s}(nj-mk)}
    [a_n,a_m^\dagger]
    \\
    &=  \frac{1}{N_s^2}
    \sum_n
    e^{i\frac{2\pi n}{N_s}(j-k)}
    \\
    &=
      \frac{1}{N_s}\delta_{jk},
\end{split}
\end{equation}
The operators $b_j^\dagger$ therefore define a set of independent bosonic modes localized on the lattice sites. The corresponding many-body basis is orthogonal,
\begin{equation}
    \braket{\{\mathbf{r}_i\}}{\{\mathbf{r}'_i\}}
    =
    \prod_i \delta_{r_i r_i'},
\end{equation}
where the states are generated by repeated application of the creation operators,
\begin{equation}
    \ket{\mathbf r}
    =
    \prod_j
    \frac{(b_j^\dagger)^{r_j}}
    {\sqrt{r_j!}}
    \ket{0}.
\end{equation}
The basis $\{\ket{\mathbf r}\}$ may therefore be interpreted as a coordinate-space occupation basis: the integer $r_j$ specifies the number of bosonic excitations localized at lattice site $j$. 
The free Hamiltonian can thus be expressed as
\begin{equation}
    H_0
    =
    \sum_{jk}
    h_{jk}\, b_j^\dagger b_k,
\end{equation}
where the hopping matrix is given by
\begin{equation}
    h_{jk}
    =
    \sum_n
    \omega_n \,
    e^{i\frac{2\pi n}{N_s}(j-k)}.
\end{equation}
In the non-relativistic regime $m \gg |k_n|$ (with $k_n \sim n/N_s$), the dispersion relation writes 
\(
\omega_n \simeq m + \frac{k_n^2}{2m}.
\)
This yields a local hopping structure. In this limit, the hopping matrix becomes approximately tridiagonal,
\begin{equation}
    h_{jk}
    =
    m N_s \delta_{jk}
    -
    \frac{N_s}{2m}
    \left(
    \delta_{j,k+1}
    +
    \delta_{j,k-1}
    \right).
\end{equation}
The interacting Hamiltonian in this basis is obtained by applying the lattice Fourier transform to the interaction expressed in the harmonic-oscillator and momentum-space representations. One then has
\begin{equation}
    H_\lambda
    =
    \frac{\lambda_0 }{4!}
    \sum_{ijkl}
    U_{ijkl}
    \Big(
    b_i b_j b_k b_l
    + 4 b_i^\dagger b_j b_k b_l
    + 6 b_i^\dagger b_j^\dagger b_k b_l
    + 4 b_i^\dagger b_j^\dagger b_k^\dagger b_l
    + b_i^\dagger b_j^\dagger b_k^\dagger b_l^\dagger
    \Big),
\end{equation}
where the tensor $U_{ijkl}$ is given by
\begin{equation}
    U_{ijkl}
    =
    \sum_{n_1+n_2+n_3+n_4=0}
    \frac{1}{\prod_{p=1}^4 \sqrt{2L\omega_{n_p}}}
    \exp\!\left[
    i\frac{2\pi}{N_s}(n_1 i + n_2 j + n_3 k + n_4 l)
    \right].
\end{equation}

Let us emphasize here that 
the tensor $U$ encodes the non-local structure of the interaction in HO${}_x$ representation.
As a consequence, the harmonic-oscillator basis in coordinate space does not simultaneously diagonalize the free and interacting parts of the Hamiltonian: unlike the momentum-space representation, the free term is not diagonal in the site basis, while the interaction is also non-local.

However, the off-diagonal structure of both $h_{jk}$ and $U_{ijkl}$ is exponentially suppressed in gapped theories. More precisely, the decay is controlled by the finite correlation length $\xi \sim 1/m_{\mathrm{gap}}$, so that matrix elements connecting sites separated by a distance $|i-j| \gg \xi$ fall off as $\sim e^{-|i-j|/\xi}$. 
As already mentioned in the Introduction,
it is precisely for these reasons that it is of interest to study quantum simulations of the $\phi^4$ model in the HO${}_x$ representation. 

%

\section{Band diagonal structure of the $\phi^4$ Hamiltonian}
\label{sec4}

In this section, we realize a numerical study of the properties of the
$\phi^4$ Hamiltonian in the 
HO${}_x$ representation.  
In particular, we analyze the sparsity and structural properties of the Hamiltonian, 
properties which follow from the corresponding one-body and two-body matrix elements, $h_{jk}$ and $U_{ijkl}$, introduced in the previous section.

\begin{figure}[H]
    \centering
    \begin{subfigure}[b]{0.48\textwidth}
        \centering
        \includegraphics[width=\textwidth]{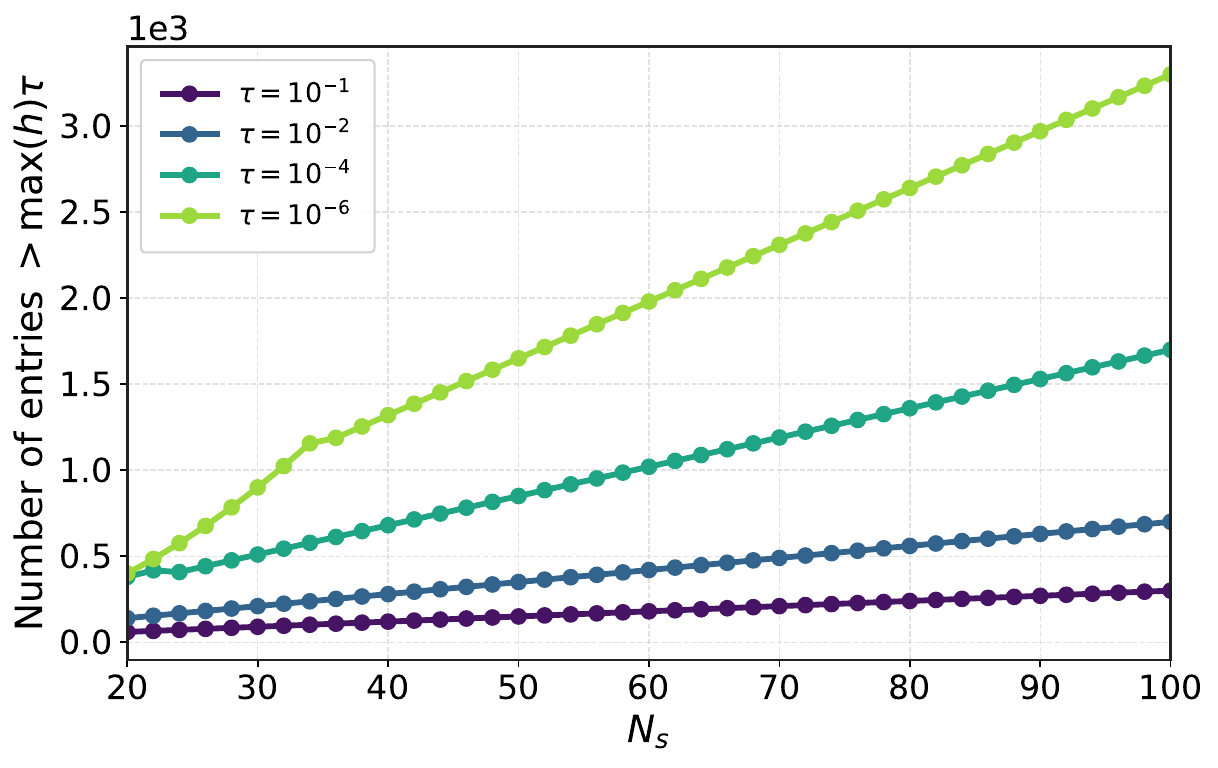}
        \caption{}
        \label{fig:left}
    \end{subfigure}
    \hfill
    \begin{subfigure}[b]{0.48\textwidth}
        \centering
        \includegraphics[width=\textwidth]{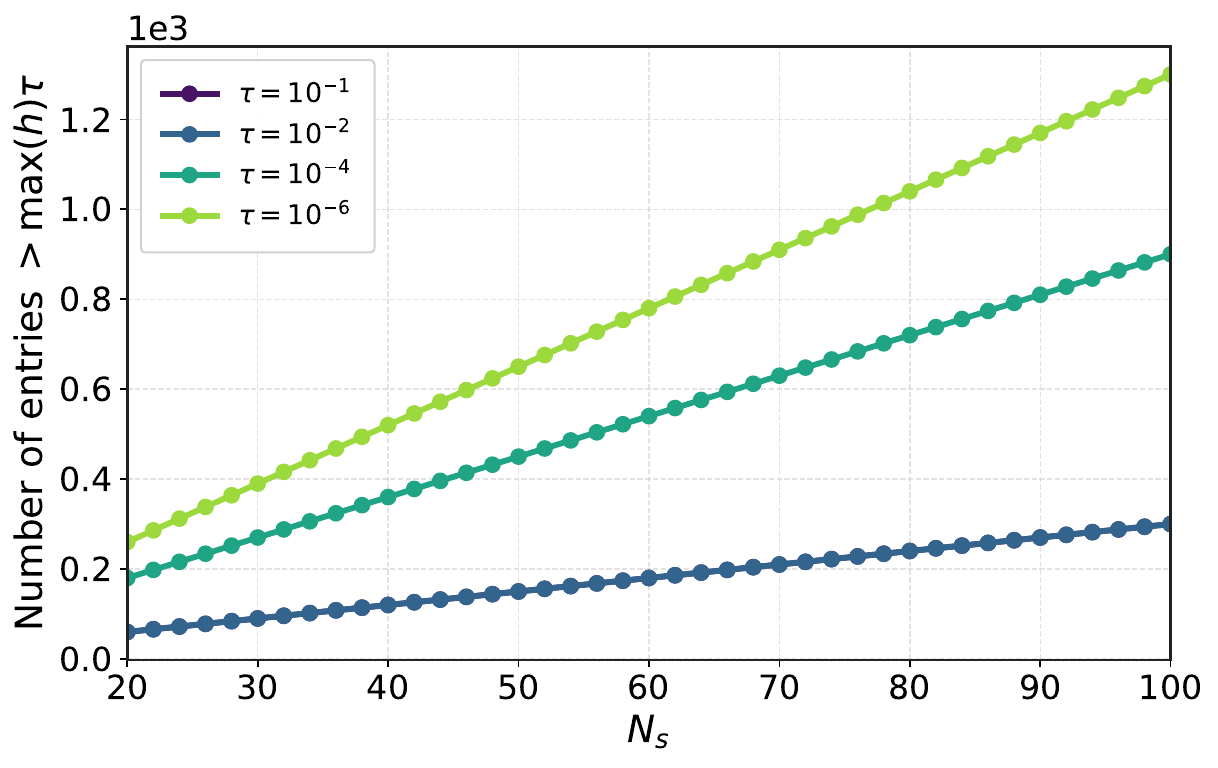}
        \caption{}
        \label{fig:right}
    \end{subfigure}
    \caption{Number of non-zero entries in the hopping matrix $h$ as a function of the lattice size $N_s$, for $m=1$ (a) and $m=3$ (b). The linear scaling demonstrates the effective band-diagonal structure of $h$. For $m=3$, the results for $\tau=10^{-2}$ and $\tau=10^{-1}$ coincide, indicating that only the tridiagonal contributions remain significant.}
    \label{fig:band_diag}
\end{figure}

Let us start this numerical analysis by investigating
the scaling of the number of non-zero entries in the hopping matrix $h_{jk}$ as a function of the system size $N_s$ and the mass $m$. Representative results for $m=1$ and $m=3$ are shown in Fig.~\ref{fig:band_diag}.

One can notice that the number of non-zero entries in $h_{jk}$ scales linearly with the number of lattice sites $N_s$. This is characteristic of a band-diagonal structure in which only couplings within a finite spatial range contribute significantly.
As a consequence, one can introduce an effective bandwidth cutoff $C_h$ such that
\begin{equation}
    |i-j| \geq C_h \;\;\Longrightarrow\;\; h_{ij} = 0.
    \label{C_T}
\end{equation}
This cutoff defines an effective interaction range for the hopping matrix. Recall that in gapped theories, the underlying matrix elements decay exponentially with distance, with a characteristic scale set by the correlation length $\xi \sim 1/m_{\mathrm{gap}}$. Consequently, the cutoff $C_h$ may be chosen independently of system size $N_s$, up to exponentially small corrections.

\begin{figure}[H]
    \centering
    \includegraphics[width=0.55\textwidth]{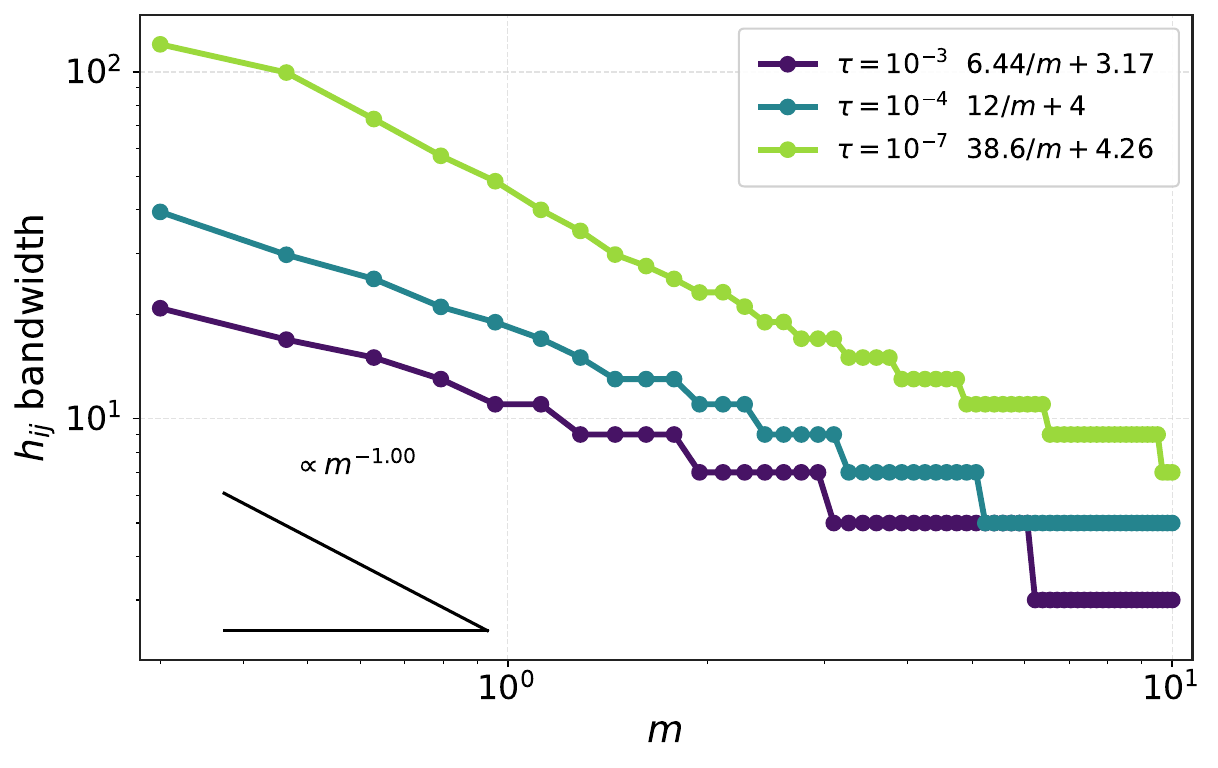}
\caption{Effective bandwidth of the hopping matrix $h_{ij}$ as a function of the mass parameter $m$. Power-law dependence of the effective bandwidth on $m$. The decreasing bandwidth indicates that the hopping matrix becomes increasingly localized as the mass increases.}
    \label{fig:width(m)}
\end{figure}

Furthermore, note that 
the number of non-zero entries is significantly reduced for $m=3$ compared to $m=1$ at fixed $N_s$.
The numerical dependence of the effective bandwidth on the mass parameter $m$ is illustrated in Fig.~\ref{fig:width(m)}.

Overall, the cost of digitizing the free Hamiltonian $H_0$ is comparable in both momentum and coordinate space representations. In momentum space, the Hamiltonian is exactly diagonal, whereas in coordinate space it becomes band-diagonal with an effective finite range. In both cases, the computational cost scales similarly with the number of lattice sites or modes, since the relevant quantity is the number of significantly contributing couplings, which grows linearly with the system size.

\medskip

\begin{figure}[H]
    \centering
    \begin{subfigure}[b]{0.48\textwidth}
        \centering
        \includegraphics[width=\textwidth]{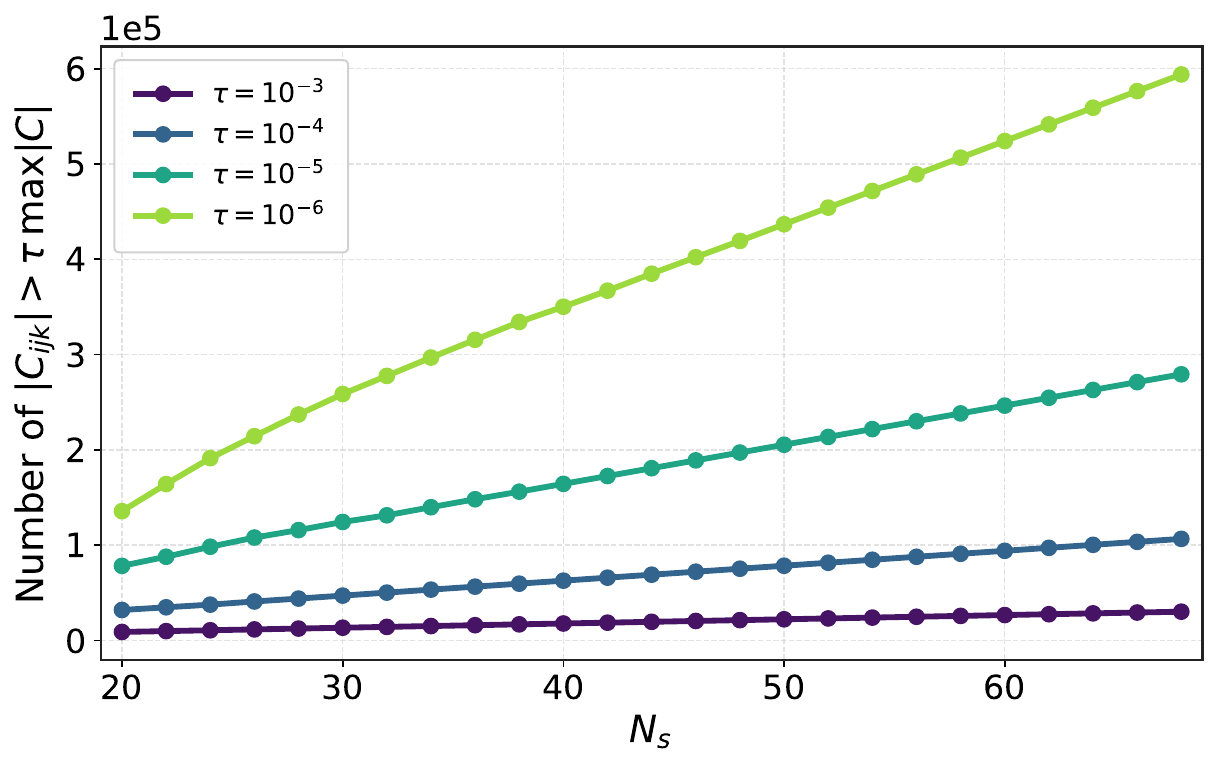}
        \caption{}
        \label{fig:left}
    \end{subfigure}
    \hfill
    \begin{subfigure}[b]{0.48\textwidth}
        \centering
        \includegraphics[width=\textwidth]{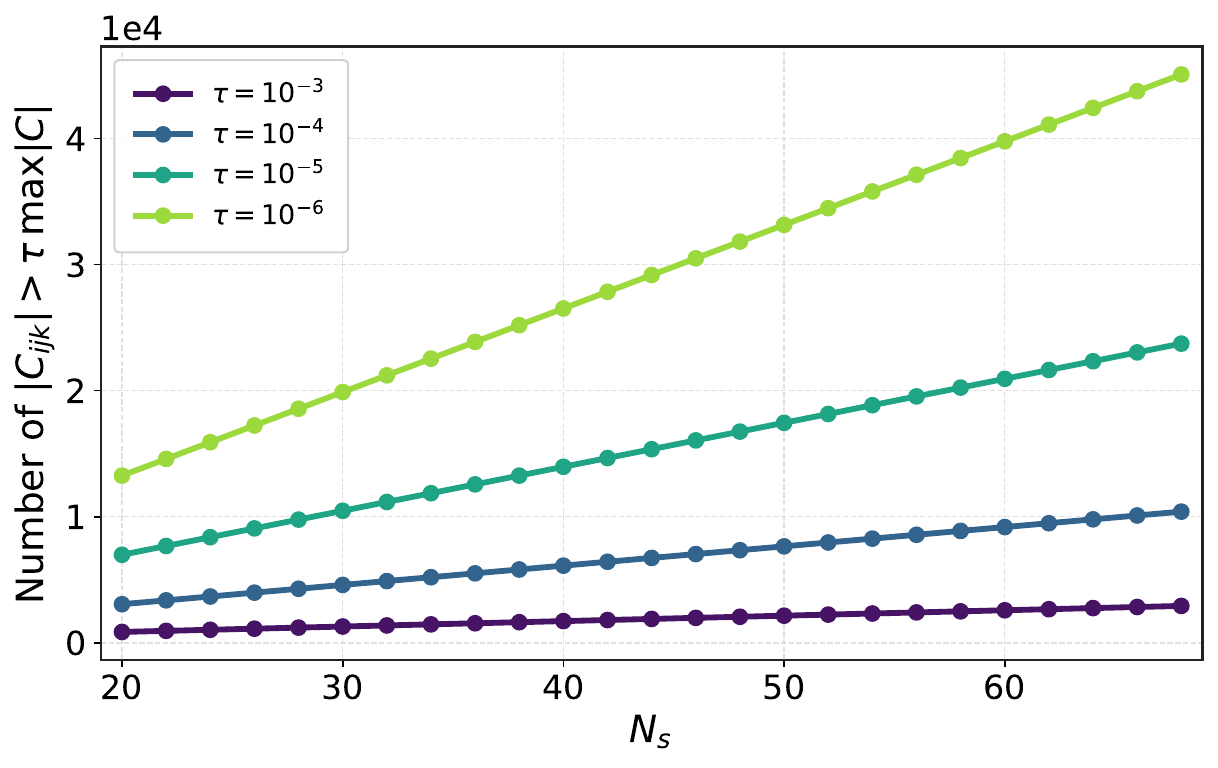}
        \caption{}
        \label{fig:right}
    \end{subfigure}
    \caption{Number of non-zero entries in the interaction tensor $U$ as a function of system size $N_s$, for $m=1$ (a) and $m=3$ (b). As for the hopping matrix, one notices a linear scaling, characteristic of a finite interaction range.}
    \label{fig:sparsity_U}
\end{figure}

On the other hand, 
the sparsity of the interaction tensor $U$ can also be quantified by measuring the number of non-zero entries as a function of the system size $N_s$. The corresponding results are shown in Fig.~\ref{fig:sparsity_U} for $m=1$ and $m=3$.

\begin{figure}[H]
    \centering
    \includegraphics[width=0.55\textwidth]{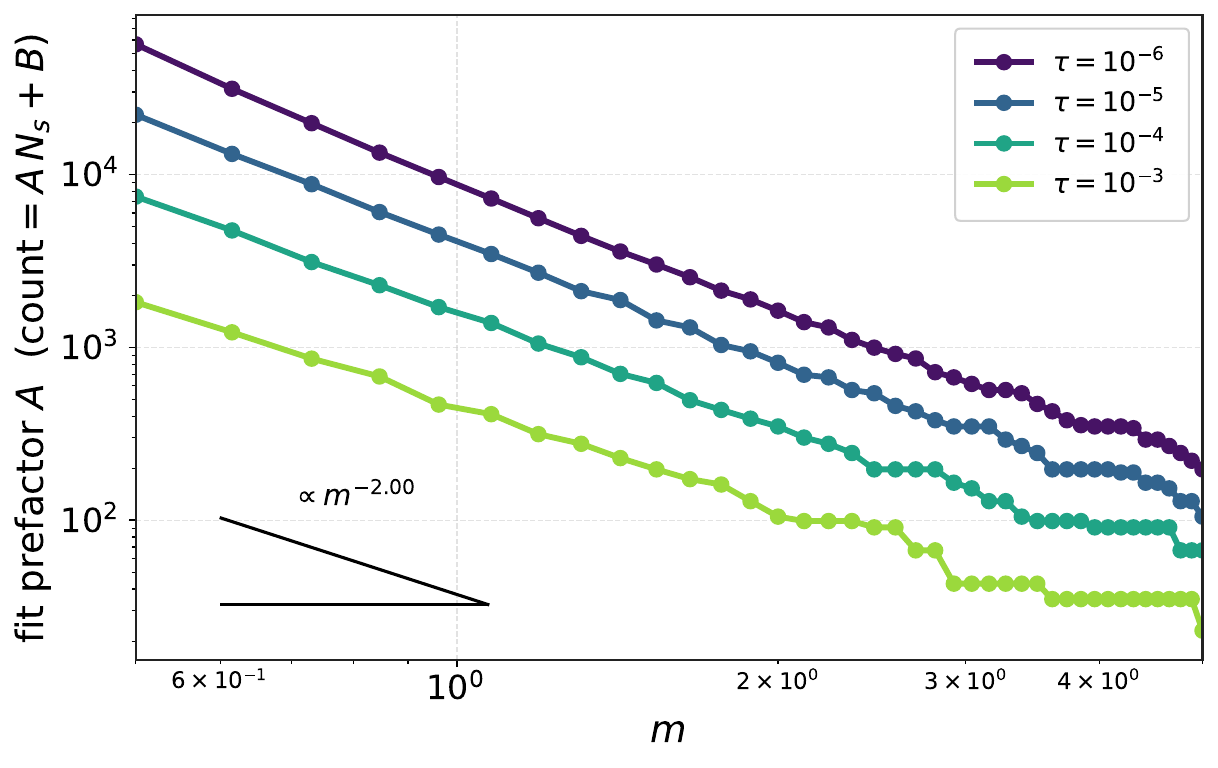}
    \caption{Effective bandwidth of $U_{ijkl}$ as a function of the mass parameter $m$. Power-law dependence of the effective bandwidth on $m$. }
    \label{fig:bandwidth_U}
\end{figure}

One can see from Fig.~\ref{fig:sparsity_U} that the number of non-zero terms scales linearly with the number of sites. 
This motivates the introduction of an effective cutoff $C_U$ which controls the range of non-local couplings. Let us define a distance function on the indices $(ijkl)$:
\begin{equation}
    d(i,j,k,l)
    =
    \max_{a,b \in \{i,j,k,l\}} |a-b|.
\end{equation}
This allows to truncate the tensor accordingly; one thus has
\begin{equation}
    d(i,j,k,l) \geq C_U \;\;\Longrightarrow\;\; U_{ijkl} = 0.
    \label{C_U}
\end{equation}
For gapped theories, the decay of these off-diagonal contributions is expected to be exponential in the separation scale, with a characteristic length controlled by the correlation length $\xi \sim 1/m_{\mathrm{gap}}$. 
As a consequence, the cutoff $C_U$ remains bounded as $N_s$ increases, up to exponentially small corrections. The dependence of the effective bandwidth on the mass is shown in Fig.~\ref{fig:bandwidth_U}.

Let us recall that, in momentum space, the number of non-zero contributions scales as $\mathcal{O}(N_s^3)$, since the interaction is non-vanishing whenever the momentum conservation constraint $n_1+n_2+n_3+n_4=0$ is satisfied. 
In contrast, we have shown numerical evidence that in coordinate space the effective number of non-zero terms grows only linearly with $N_s$ under the locality-based truncation. This reduction in the number of terms thus provides a computational advantage for quantum simulations in the HO${}_x$ representation studied in this paper.

\section{Ground state energy and mass gap}

In this section, we compare the momentum-space and coordinate-space representations within the harmonic-oscillator (HO) basis. In particular, we verify that numerical diagonalization of the Hamiltonian in both representations yields asymptotically consistent values for the ground state energy $E_0$ and the mass gap $m_{\mathrm{ph}} = E_1 - E_0$, up to finite-size and truncation effects.

Recall that a numerical implementation requires a truncation of the local Hilbert space. While the number of lattice sites (or momentum modes) is finite due to the ultraviolet cutoff $a^{-1}$ and infrared cutoff $N_s$ (with $j \in [1,N_s]$ and $n \in [-(N_s-1)/2, (N_s-1)/2]$), each site still carries an infinite-dimensional bosonic degree of freedom. We therefore impose a cutoff $N_\phi-1$ on the local occupation number, restricting the basis to states with $r_j \in \{0,1,\ldots,N_\phi-1\}$.

We compare our approach with the Hamiltonian truncation method of ~\cite{Rychkov_2015}. In that framework, the Fock space is constructed explicitly and restricted to states with total energy below a hard cutoff $\Lambda$. Projecting the Hamiltonian onto this truncated space enables the computation of low-energy eigenstates via numerical diagonalization. Implementing such sharp energy truncation on a quantum computer is hard because it requires to go through all the Fock states with an energy below $\Lambda$. This requires to loop over the whole truncated Hilbert space, which has exponential size in $\Lambda$. 
The Hamiltonian in the HO${}_p$ representation, with cutoffs $(N_\max, N_\phi)$ can instead be directly encoded. 
In the following, we denote by $H_\Lambda$ the momentum-space Hamiltonian equipped with a sharp energy cutoff, by $H_x$ the momentum-space Hamiltonian truncated via finite mode and local occupation cutoffs $(N_{\max}, N_\phi)$, and by $H_p$ the coordinate-space Hamiltonian defined on a lattice with parameters $(N_s, N_\phi)$.

Our analysis proceeds in two steps:
First, we verify that numerical diagonalization of $H_\Lambda$ and $H_p$ yields asymptotically equivalent results for low-energy observables. We then compare the spectra of $H_p$ and $H_x$ across different parameter regimes.

\begin{figure}[H]
    \centering
    \includegraphics[width=0.56\linewidth]{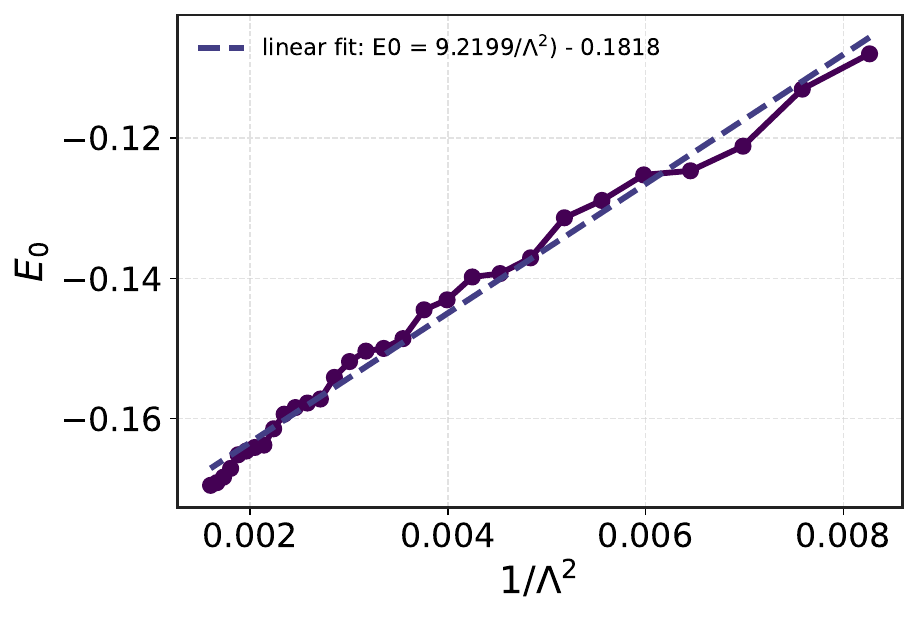}
    \caption{Ground-state energy obtained from $H_\Lambda$ as a function of $1/\Lambda^2$ for $L=5$. The approximately linear behavior enables an extrapolation to the infinite-cutoff limit. The extrapolated energy at infinite cutoff is $E_0=-0.1818$. }
    \label{fig:E0Hlam}
\end{figure}

\begin{figure}[H]
    \centering
    \includegraphics[width=\textwidth]{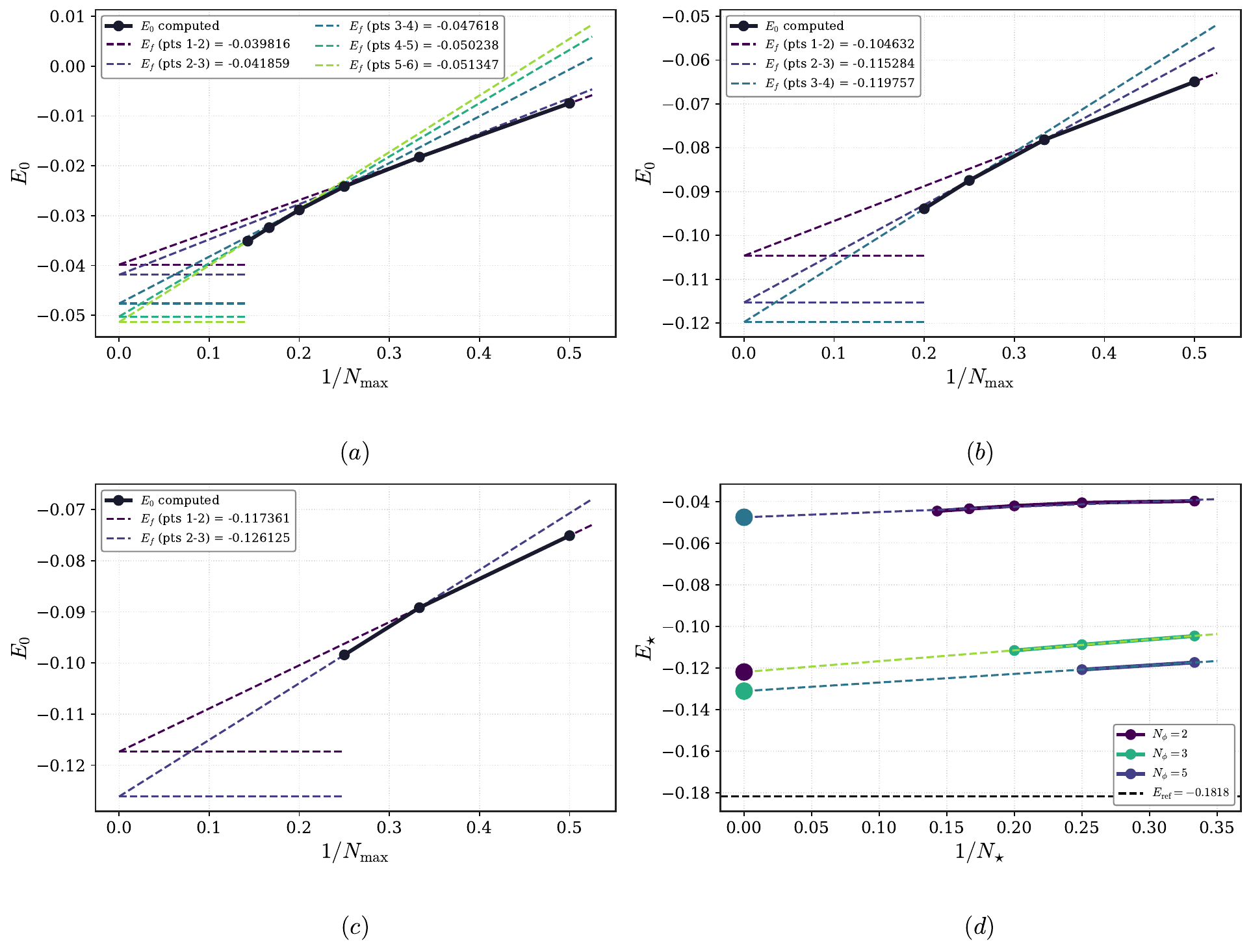}
    \caption{Convergence of the ground-state energy obtained from $H_p$ with respect to the momentum cutoff $N_{\mathrm{max}}$ and the local Hilbert space dimension $N_\phi$.  \textbf{(a)--(c)} Ground-state energy as a function of the inverse momentum cutoff $1/N_{\mathrm{max}}$ for $N_\phi=2$, $3$, and $5$. \textbf{(d)} Extrapolation of the converged ground-state energy to the $N_{\mathrm{max}}\rightarrow\infty$ limit for increasing $N_\phi$. }
    \label{fig:E0_conv}
\end{figure}

We begin by comparing the ground-state energy obtained from the numerical diagonalization of $H_\Lambda$ and $H_p$. 
For the sharp cutoff approach, the ground-state energy is expected to converge as $\mathcal{O}(\Lambda^{-2})$ (see again \cite{Rychkov_2015}). Figure~\ref{fig:E0Hlam} exhibits $E_0$ as a function of $1/\Lambda^2$, from which we extrapolate a ground-state energy of $E_0=-0.1818$ for a system of size $L=5$. Let us mention that we use the raw truncation and that subleading contribution in $1/\Lambda$ are not included here.
We next consider the momentum-space Hamiltonian $H_p$, in which the Hilbert space is truncated by imposing cutoffs on both the maximum momentum mode $N_{\mathrm{max}}$ and the local occupation number $N_\phi$. Figure \ref{fig:E0_conv} displays the ground-state energy as a function of $1/N_{\mathrm{max}}$ for several values of $N_\phi$. As expected, increasing either cutoff systematically improves the convergence toward the asymptotic value. 

For sufficiently large values of $N_{\mathrm{max}}$ and $N_\phi$, the extrapolated ground-state energy converges towards with the value obtained using Hamiltonian truncation. This agreement confirms that the low-energy spectrum is insensitive to the particular ultraviolet regularization employed, provided that the truncation parameters are chosen sufficiently large.
The convergence obtained with the $(N_{\max},N_\phi)$ truncation, however, comes at a higher computational cost. The sharp energy cutoff retains only the Fock states that contribute directly to the low-energy sector, whereas imposing independent cutoffs on the momentum modes and local occupation numbers generates a much larger Hilbert space containing many high-energy states that do not play an important role in the low-energy spectrum. 
Furthermore, note that in ~\cite{Rychkov_2015}, S. Rychkov and L. Vitale use the continuum dispersion relation, while we stick here with the lattice dispersion relation \eqref{omega}. This is another source of discrepancy that disappears in the large cutoff limit. 

\begin{figure}[H]
    \centering

    \begin{subfigure}[b]{0.48\textwidth}
        \centering
        \includegraphics[width=\linewidth]{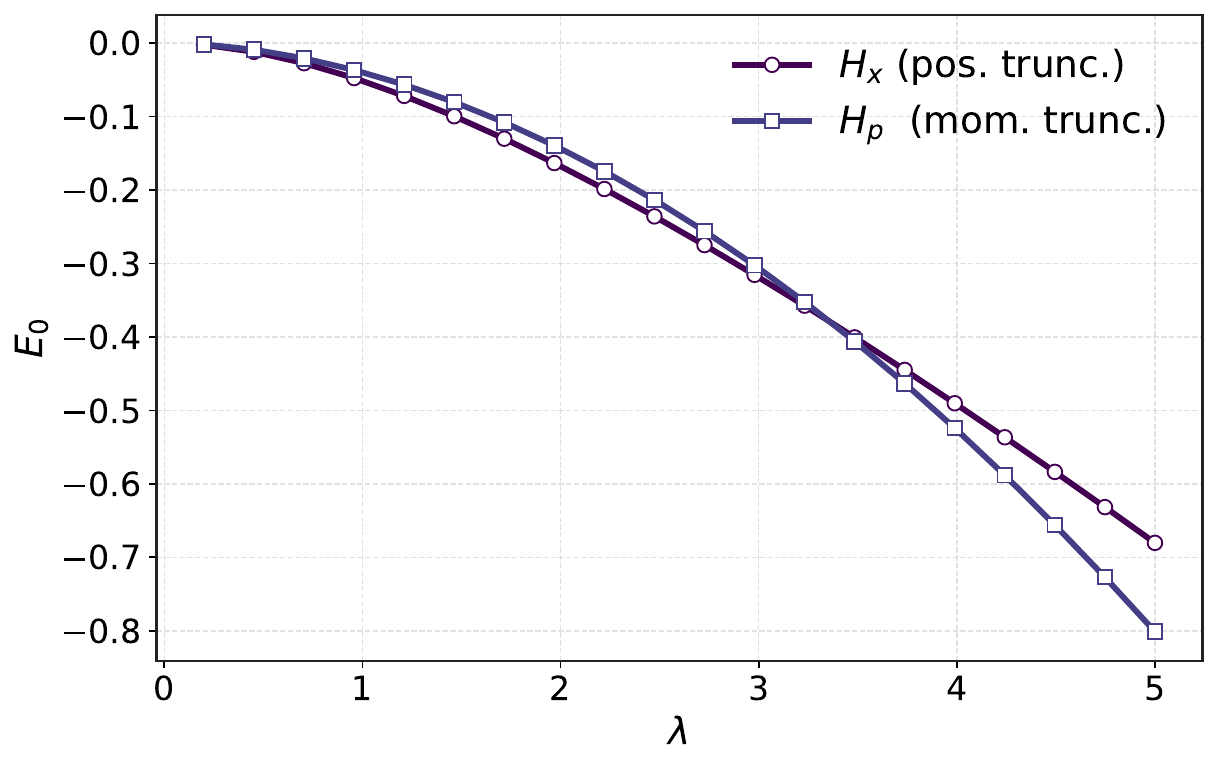}
        \caption{}
        \label{fig:E0_pos_vs_en}
    \end{subfigure}
    \hfill
    \begin{subfigure}[b]{0.48\textwidth}
        \centering
        \includegraphics[width=\linewidth]{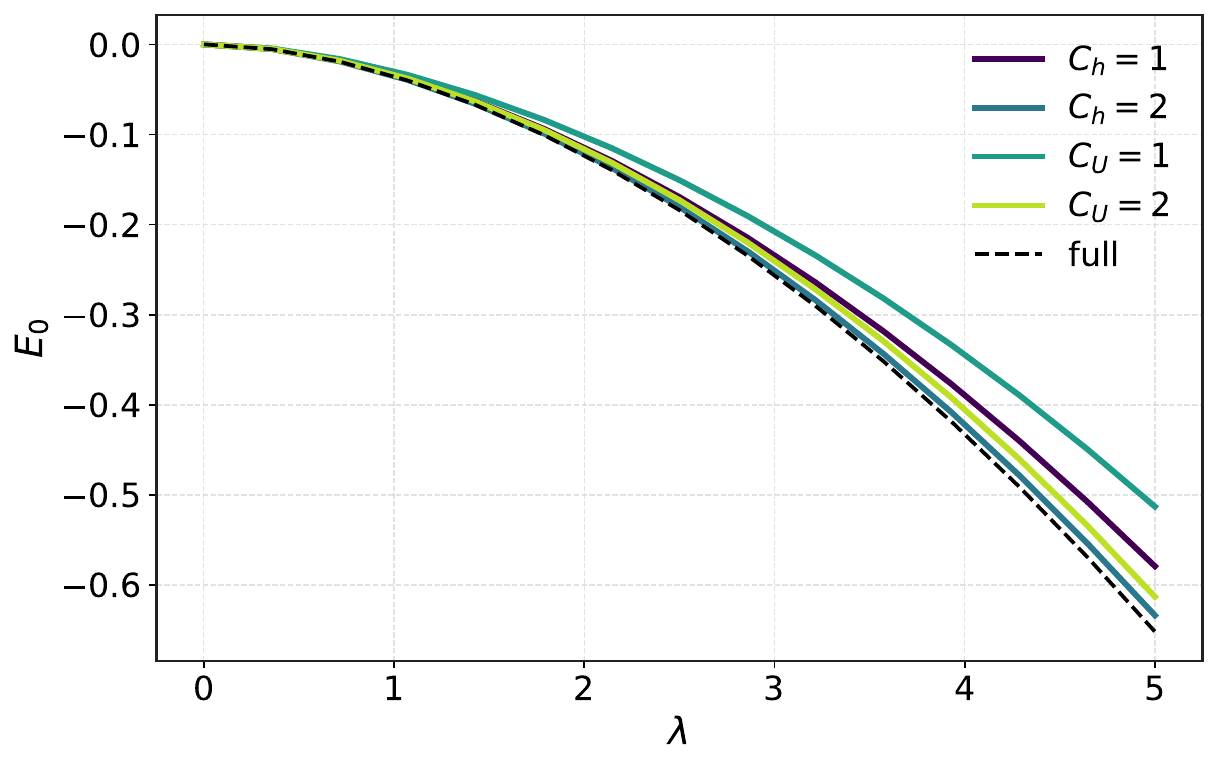}
        \caption{}
        \label{fig:CTCUeffect}
    \end{subfigure}

    \caption{Comparison of the low-energy spectra. \textbf{(a)} Ground-state energy as a function of the coupling $\lambda$ obtained by exact diagonalization of the momentum-space Hamiltonian $H_p$ and the coordinate-space Hamiltonian $H_x$ for $L=5$, $N_s=5$, and $N_\phi=5$. \textbf{(b)} Ground-state energy as a function of $\lambda$ for different hopping and interaction bandwidth cutoffs. Small effective bandwidths are sufficient to accurately reproduce the low-energy spectrum.}
    \label{fig:spectrum_comparison}
\end{figure}

Let us now compare the spectra obtained from the Hamiltonians $H_p$ and $H_x$. Since these two Hamiltonians are related by a change of basis, they are expected to yield identical physical observables in the limit of large truncation parameters. We also investigate the effect of the bandwidth cutoffs introduced in eqs.~\eqref{C_T} and \eqref{C_U} on the low-energy spectrum. The corresponding results are summarized in Fig.~\ref{fig:spectrum_comparison}.

Figure~\ref{fig:spectrum_comparison}(a) shows the ground-state energy as a function of the coupling $\lambda$. We observe excellent agreement between the momentum- and coordinate-space formulations, confirming that both representations yield the same low-energy physics within the considered truncation.

Figure~\ref{fig:spectrum_comparison}(b) illustrates the dependence of the ground-state energy on the hopping and interaction bandwidths for $N_\phi=6$ and $N_s=6$. We find that retaining only nearest-neighbor hopping ($C_h=1$) together with an interaction bandwidth $C_U=2$ is sufficient to reproduce the ground-state energy with essentially the same accuracy as that obtained from the full Hamiltonian. This confirms that the dominant contributions to both the free and interacting parts of the Hamiltonian are localized in the harmonic-oscillator coordinate-space basis. In particular, the free Hamiltonian is already well approximated by its tridiagonal form, while the exponentially suppressed off-diagonal elements of $U_{ijkl}$ have a negligible effect on the low-energy spectrum.
The results shown in Fig.~\ref{fig:spectrum_comparison}(b) correspond to a bare mass $m=2$. As discussed in Sec.~\ref{sec4}, the effective bandwidths of both $h_{jk}$ and $U_{ijkl}$ decrease as the mass increases. The truncated Hamiltonian is therefore expected to become even more accurate in the large-mass regime.

As a final comparison, we examine the mass gap,
\begin{equation}
    m_{\mathrm{ph}} = E_1 - E_0,
\end{equation}
computed in the HO$_x$ representation for different cutoff values $C_h$ and $C_U$. Near the critical point, the mass gap is expected to scale as
\begin{equation}
    m_{\mathrm{ph}} = C |\lambda - \lambda_c|^\nu,
\end{equation}
which provides a way to extract the critical coupling $\lambda_c$. A useful benchmark is provided by Ref.~\cite{Rychkov_2015}, where the critical coupling was estimated to be $\lambda_c \simeq 2.97(14)$.

\begin{figure}[H]
    \centering
    \includegraphics[width=0.60\linewidth]{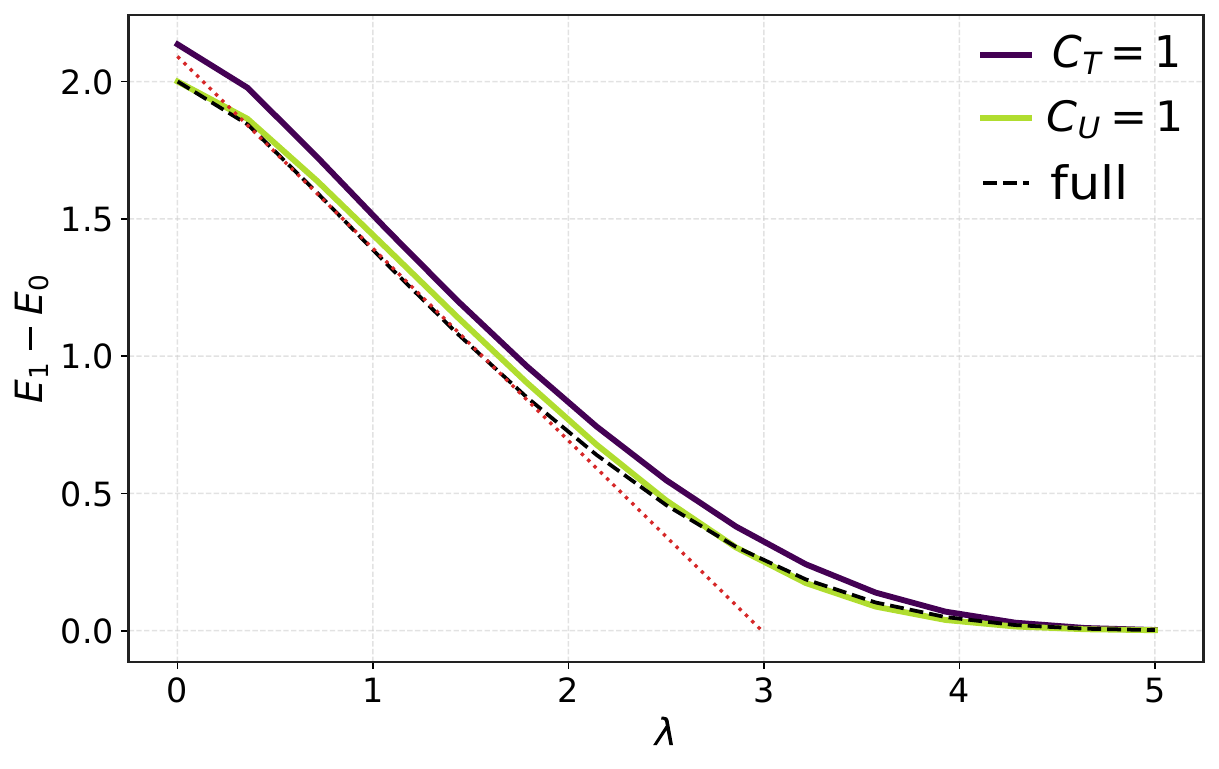}
    \caption{Mass gap as a function of the coupling $\lambda$ for different bandwidth cutoffs $C_h$ and $C_U$. A fit to the approximately linear region of the untruncated result yields a critical coupling $\lambda_c \simeq 3$, in agreement with Ref.~\cite{Rychkov_2015}. The bandwidth cutoffs introduce only minor deviations from the full Hamiltonian.}
    \label{fig:mph}
\end{figure}
Figure~\ref{fig:mph} shows that the bandwidth cutoffs have only a minor effect on the mass gap over the range of couplings considered. Fitting the approximately linear region of the curve without cutoffs yields $\lambda_c \simeq 3$, in good agreement with the value reported in Ref.~\cite{Rychkov_2015}. This further demonstrates that a strongly banded approximation of the Hamiltonian accurately reproduces not only the ground-state energy but also the low-energy excitation spectrum.

\section{Qubit encodings and implementation cost}
\label{sec5}
In this section, we estimate the resources required to implement the Hamiltonian on a quantum computer. We first review the unary and binary boson-to-qubit encodings before deriving estimates for the number of qubits, the number of Pauli strings, and the Pauli $1$-norm of the resulting Hamiltonian.
\subsection{Unary and binary encoding}

In both the HO${}_p$ and HO${}_x$ representations, the annihilation operator is given by
\begin{equation}
    a_n
    =
    \sum_{r=0}^{N_\phi-1}
    \sqrt{r+1}\,
    \ket{r}_n \bra{r+1}_n,
\end{equation}
where $n$ labels a momentum mode in the HO${}_p$ representation and a lattice site in the HO${}_x$ representation.

To implement the Hamiltonian on a quantum computer, the bosonic creation and annihilation operators must be mapped onto qubit operators. The two most common boson-to-qubit mappings are the binary and unary encodings (see \cite{unary, binary, bosonic_mapping_1}). For the sake of completeness, we briefly review both encodings below.

\paragraph{Binary encoding}

An efficient encoding of bosonic degrees of freedom onto qubits is obtained through the binary representation of the occupation number associated with each mode. This encoding requires $n_q=\log_2(N_\phi)$ qubits.
The occupation number $r$ is represented through its binary decomposition,
\begin{equation}
    r
    =
    \sum_{l=0}^{n_q-1}
    2^l b_l,
\end{equation}
with $b_l \in \{0,1\}$. The corresponding Fock state is encoded as
\begin{equation}
    \ket{r}_n
    =
    \bigotimes_{l=0}^{n_q-1}
    \ket{b_l}_{n_q n+l}.
\end{equation}
This encoding achieves an exponential compression of the local Hilbert space, reducing the number of qubits per mode/site from $\mathcal{O}(N_\phi)$ to $\mathcal{O}(\log N_\phi)$.
Within this representation, the transition operator $\ket{r}_n\bra{r+1}_n$ factorizes as
\begin{equation}
    \ket{r}_n\bra{r+1}_n
    =
    \bigotimes_{l=0}^{n_q-1}
    \ket{b_l}_{n_q n+l}
    \bra{b_l'}_{n_q n+l},
\end{equation}
where $\{b_l\}$ and $\{b_l'\}$ denote the binary coefficients of $r$ and $r+1$, respectively. This decomposition allows the ladder operators to be expressed as sums of tensor products of single-qubit operators, ultimately yielding a Pauli-string representation of the Hamiltonian. Each single-qubit operator $\ket{b}\bra{b'}$ can be expressed in terms of Pauli matrices according to
\begin{align*}
    \ket{0}\bra{0}
    &=
    \frac{I+\sigma^z}{2},
    &
    \ket{1}\bra{0}
    &=
    \frac{\sigma^x-i\sigma^y}{2},
    \\
    \ket{0}\bra{1}
    &=
    \frac{\sigma^x+i\sigma^y}{2},
    &
    \ket{1}\bra{1}
    &=
    \frac{I-\sigma^z}{2}.
\end{align*}
Consequently, the creation and annihilation operators can be expressed as linear combinations of Pauli strings. To illustrate the construction, let us consider the cases $N_\phi=2$ and $N_\phi=4$.

For $N_\phi=2$, a single qubit is sufficient to encode each mode and the annihilation operator reduces to
\begin{equation}
    a_n^{(N_\phi=2)}
    =
    (\ket{0}\bra{1})_n
    =
    \sigma_n^+.
\end{equation}

For $N_\phi=4$, two qubits are required per mode and one obtains
\begin{equation}
\begin{split}
    a_n^{(N_\phi=4)}
    &=
    \ket{0}\bra{1}
    +
    \sqrt{2}\ket{1}\bra{2}
    +
    \sqrt{3}\ket{2}\bra{3}
    \\
    &=
    \ket{00}\bra{01}
    +
    \sqrt{2}\ket{01}\bra{10}
    +
    \sqrt{3}\ket{10}\bra{11}.
\end{split}
\end{equation}
Each transition operator can then be factorized into tensor products of single-qubit operators and subsequently expanded into Pauli strings using the identities above.

More generally, the annihilation operator admits a decomposition of the form
\begin{equation}
    a_n
    =
    \sum_{l=1}^{\lambda}
    \alpha_l P_l,
\end{equation}
where the $P_l$ are Pauli strings and $\alpha_l \in \mathbb{C}$. Since each qubit contributes at most four possible single-qubit operators $\{I,\sigma^x,\sigma^y,\sigma^z\}$, the number of terms satisfies
\begin{equation}
    \lambda
    \leq
    4^{n_q} = N_\phi^2,
\end{equation}
Each Pauli string acts non-trivially on at most $n_q$ qubits. Applying the same procedure to all terms in the Hamiltonian yields a Pauli decomposition
\begin{equation}
    H
    =
    \sum_{l=1}^{\Lambda}
    \alpha'_l P_l.
\end{equation}

\paragraph{Unary encoding}
Another mapping of bosonic degrees of freedom onto qubits is provided by the unary encoding. In contrast to the binary encoding, each occupation number is represented explicitly, requiring $n_q = N_\phi$ qubits per mode/site.
The encoding is defined by
\begin{equation}
    \ket{r}_n
    =
    \ket{1}_{n_q n + r},
\end{equation}
with all remaining qubits associated with the mode in the $\ket{0}$ state. In other words, the occupation number is encoded in the position of a single excitation among the $n_q$ qubits assigned to the mode. The $n$-th mode/site has occupation number $r$ if and only if the qubit indexed by $n_q n+r$ is in the state $\ket{1}$. Within this representation, transitions between neighboring occupation levels take a particularly simple form,
\begin{equation}
    \ket{r}_n\bra{r+1}_n
    =
    \sigma^+_{n_q n+r}\,
    \sigma^-_{n_q n+r+1}.
\end{equation}
Indeed,
\begin{equation}
\begin{split}
    \sigma^+_{n_q n+r}
    \sigma^-_{n_q n+r+1}
    \ket{r+1}_n
    &=
    \sigma^+_{n_q n+r}
    \sigma^-_{n_q n+r+1}
    \ket{1}_{n_q n+r+1}
    \\
    &=
    \ket{1}_{n_q n+r}
    =
    \ket{r}_n,
\end{split}
\end{equation}
while all other occupation states are annihilated. Consequently, the bosonic annihilation operator can be written as
\begin{equation}
\begin{split}
    a_n
    &=
    \sum_{r=0}^{N_\phi-1}
    \sqrt{r+1}\,
    \sigma^+_{n_q n+r}
    \sigma^-_{n_q n+r+1}
    \\
    &=
    \sum_{l=1}^{\lambda}
    \alpha_l P_l,
\end{split}
\end{equation}
with
\begin{equation}
    \lambda = N_\phi,
\end{equation}
and where each Pauli string acts on exactly two qubits $|P_l| = 2$.
Compared with the binary encoding, the unary representation requires a larger number of qubits but yields substantially simpler operator expressions. In particular, the number of Pauli strings appearing in a ladder operator scales only linearly with the local occupation cutoff.

The Hamiltonian can again be expressed as a sum of Pauli strings,
\begin{equation}
    H
    =
    \sum_{l=1}^{\Lambda}
    \alpha'_l P_l.
    \label{Paulistrings}
\end{equation}

\subsection{Cost estimation}

We estimate the number of Pauli strings for the Hamiltonian in the HO${}_p$ and HO${}_x$ representations, in both binary and unary encoding. 

For the Hamiltonian in the HO${}_p$ representation,  
the quartic interaction couples up to four bosonic modes simultaneously.
In the binary encoding, the quadratic Hamiltonian generates Pauli strings acting on at most $2n_q$ qubits, while the interaction term generates Pauli strings acting on at most $4n_q$ qubits. The number of possible Pauli strings on these supports is therefore bounded by $4^{2n_q}=N_\phi^4$ and $4^{4n_q}=N_\phi^8$, respectively, yielding the estimate
\begin{equation}
    \Lambda
    \leq
    (2N_{\max}+1)N_\phi^4+ (2N_{\max}+1)^3N_\phi^8.
\end{equation}
In unary encoding, the free and interacting contributions generate at most
\begin{equation}
    \Lambda
    \leq
    (2N_{\mathrm{max}}+1)N_\phi^2
    +    (2N_{\mathrm{max}}+1)^3 N_\phi^4
\end{equation}
terms. Furthermore, each Pauli string satisfies
\begin{equation}
    |P_l| \leq 8,
\end{equation}
since the quartic interaction involves at most four ladder operators, each contributing support on two qubits.

For the Hamiltonian in the HO${}_x$ representation, since the bosonic operators are encoded in exactly the same manner as in the momentum-space formulation, the complexity of the resulting Pauli decomposition is determined by the structure of the coefficients $h_{jk}$ and $U_{ijkl}$.
The quadratic contribution
\begin{equation}
    H_0
    =
    \sum_{jk}
    h_{jk}
    a_j^\dagger a_k
\end{equation}
contains at most $N_s^2$ operator pairs $a_j^\dagger a_k$.
In unary encoding, each ladder operator expands into $N_\phi$ Pauli strings, the total number of Pauli strings required to represent $H_0$ is bounded by
\begin{equation}
    \Lambda
    \lesssim
    N_s^2 N_\phi^2.
\end{equation}
Similarly, the quartic interaction term contains at most $N_s^4$ combinations of site indices, yielding the estimate
\begin{equation}
    \Lambda
    \lesssim
    N_s^4 N_\phi^4.
\end{equation}
However, as shown in Sec.~\ref{sec4}, both the hopping matrix $h_{jk}$ and the interaction tensor $U_{ijkl}$ exhibit an effective locality structure, with matrix elements that decay exponentially away from the diagonal. One may therefore truncate the Hamiltonian by retaining only couplings within bandwidths $C_h$ and $C_U$.

After this truncation, the number of retained coefficients scales only linearly with the system size,
\begin{equation}
    N_h \sim C_h N_s,
    \qquad
    N_U \sim C_U N_s,
\end{equation}
leading to the estimate
\begin{equation}
    \Lambda
    \lesssim
    C_h N_s N_\phi^2
    +
    C_U N_s N_\phi^4.
\end{equation}
The resulting Pauli decomposition therefore grows linearly with the number of lattice sites, rather than quadratically or quartically, making the coordinate-space representation particularly attractive for large systems.

For the binary encoding, each bosonic ladder operator decomposes into at most $\mathcal{O}(N_\phi^2)$ Pauli strings. Repeating the same counting argument yields
\begin{equation}
    \Lambda
    \lesssim
    C_h N_s N_\phi^4
    +
    C_U N_s N_\phi^8,
\end{equation}
up to numerical prefactors and possible reductions arising from symmetries or cancellations in the Pauli expansion.

We now compare the cost of implementing the momentum- and position-space Hamiltonians using both unary and binary encodings. We consider the Pauli $1$-norm of the Hamiltonian. Writing the Hamiltonian as a linear combination of Pauli strings,
\begin{equation}
    H=\sum_l \alpha_l P_l,
\end{equation}
the Pauli $1$-norm is defined as
\begin{equation}
    \|H\|_1=\sum_l |\alpha_l|.
\end{equation}
This quantity plays a central role in several algorithms such as LCU or block encoding, where it directly enters the complexity estimates.
As a representative example, we consider an encoding using two qubits per site (or mode), corresponding to a local Hilbert space of dimension four. In this case, the annihilation operator admits a particularly simple representation in the unary encoding,
\begin{equation}
    a_n
    =
    \sigma^+_{2n}\sigma^-_{2n+1},
\end{equation}
which consists of a single Pauli string.
By contrast, the binary encoding yields a more involved decomposition,
\begin{equation}
\begin{split}
    a_n
    &=
    \left(
    \frac{1}{2}(1-\sqrt{3})Z
    +
    \sqrt{2}\,\sigma^+
    +
    \frac{1}{2}(1+\sqrt{3})I
    \right)_{2n}
    \\
    &\quad\otimes
    \left(
    (1+\sqrt{2}+\sqrt{3})X
    +
    (1-\sqrt{2}+\sqrt{3})Y
    \right)_{2n+1},
\end{split}
\end{equation}
which expands into several Pauli strings. Consequently, one expects the binary encoding to produce larger Pauli $1$-norms than the unary encoding, despite requiring fewer qubits. The choice of $H_p$ or $H_x$ affects only the structure of the Hamiltonian coefficients, not the local qubit representation of the bosonic operators. Consequently, the same local operator decompositions are used in both momentum and coordinate space.

\begin{figure}[H]
    \centering
    \begin{subfigure}[b]{0.48\textwidth}
        \centering
        \includegraphics[width=\textwidth]{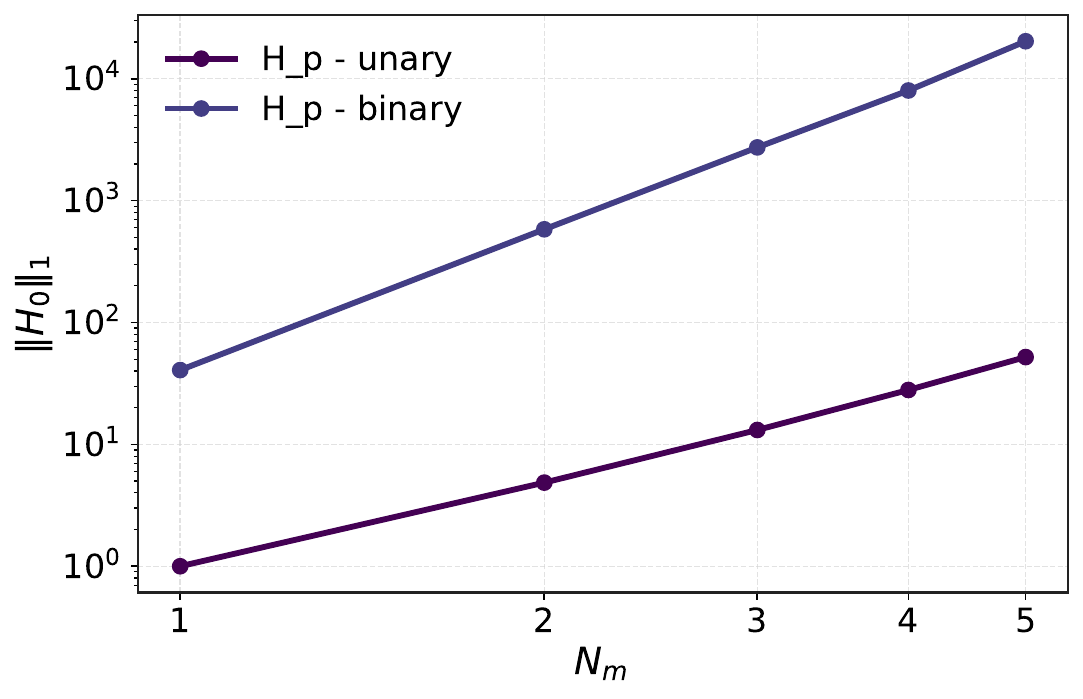}
        \caption{}
        \label{fig:1norm_encoding}
    \end{subfigure}
    \hfill
    \begin{subfigure}[b]{0.48\textwidth}
        \centering
        \includegraphics[width=\textwidth]{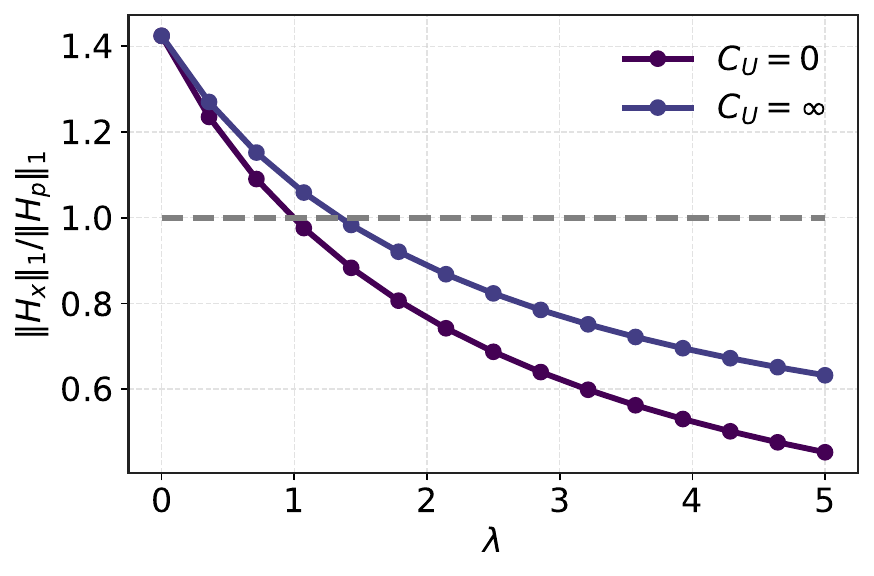}
        \caption{}
        \label{fig:1norm_ratio}
    \end{subfigure}
    \caption{Comparison of the Pauli $1$-norm for different boson-to-qubit encodings and Hamiltonian representations. \textbf{(a)} Pauli $1$-norm of the momentum-space Hamiltonian $H_p$ for unary and binary encodings. The binary encoding yields a Pauli $1$-norm approximately two orders of magnitude larger than the unary encoding. \textbf{(b)} Ratio $\|H_x\|_1/\|H_p\|_1$ as a function of the coupling strength $\lambda$ ($m=1$). The ratio decreases as the interaction becomes more important. Imposing the interaction bandwidth cutoff $C_U$ further reduces the Pauli $1$-norm of the coordinate-space Hamiltonian while having only a minor effect on the low-energy spectrum (see Fig.~\ref{fig:CTCUeffect}).}
    \label{fig:1norm}
\end{figure}

We first compare the unary and binary encodings by evaluating the Pauli $1$-norm of the momentum-space Hamiltonian $H_p$. The results are shown in Fig.~\ref{fig:1norm}(a). The binary encoding yields a Pauli $1$-norm approximately two orders of magnitude larger than the unary encoding, reflecting the larger number of Pauli strings generated by the decomposition of the bosonic ladder operators. The same trend is observed for the coordinate-space Hamiltonian, indicating that this difference originates from the encoding rather than from the choice of representation.

We next compare the momentum- and coordinate-space formulations. Figure~\ref{fig:1norm}(b) shows the ratio
\begin{equation}
    \frac{\|H_x\|_1}{\|H_p\|_1},
\end{equation}
which provides a direct comparison of the implementation cost of the two representations. In the weak-coupling regime ($\lambda \ll 1$), the free Hamiltonian dominates and the momentum-space representation has the smaller Pauli $1$-norm, since its quadratic part is diagonal. As the coupling strength increases, the interaction term becomes more important. In the coordinate-space representation, the interaction tensor is effectively band-diagonal (Sec.~\ref{sec4}), leading to fewer non-zero Pauli terms and a smaller Pauli $1$-norm. Consequently, the ratio $\|H_x\|_1/\|H_p\|_1$ decreases with increasing $\lambda$.

Figure~\ref{fig:1norm}(b) also illustrates the effect of imposing the interaction bandwidth cutoff $C_U$. Since the discarded matrix elements are exponentially suppressed, introducing this cutoff further reduces the Pauli $1$-norm while having only a minor effect on the low-energy spectrum. In practice, choosing $C_U \geq 2$ provides a good compromise between implementation cost and accuracy (see Fig.~\ref{fig:CTCUeffect}).

Overall, these results indicate that the momentum-space representation is more favorable in the weak-coupling regime, whereas the coordinate-space representation becomes more efficient as the interaction strength increases.

We conclude this section by summarizing the resource requirements for the momentum-space and coordinate-space Hamiltonians, $H_p$ and $H_x$. For fixed values of the number of lattice sites $N_s$, the momentum cutoff $N_{\max}$, and the local Hilbert-space dimension $N_\phi$, Table~\ref{tab:sumup} reports the total number of qubits $N_q$, the number of Pauli strings appearing in the Hamiltonian, for both the unary and binary encodings.
\begin{table}[h]
\centering
\begin{adjustbox}{max width=\linewidth}
\begin{tabular}{|l|cc|cc|}
\hline
& \multicolumn{2}{c|}{\textbf{Binary encoding}} &
\multicolumn{2}{c|}{\textbf{Unary encoding}} \\
\cline{2-5}
& \textbf{$H_p$-binary} &
\textbf{$H_x$-binary} &
\textbf{$H_p$-unary} &
\textbf{$H_x$-unary} \\
\hline
$N_q$ &
$(2N_\mathrm{max}+1)\log_2(N_\phi)$ &
$N_s\log_2(N_\phi)$ &
$(2N_\mathrm{max}+1)N_\phi$ &
$N_sN_\phi$ \\
\hline
$N_\mathrm{Paulis}$ &
$\lesssim(2N_\mathrm{max}+1)^4N_\phi^8$ &
$\lesssim C_hN_sN_\phi^4 + C_UN_sN_\phi^8$ &
$\lesssim (2N_\mathrm{max}+1)^4N_\phi^4$ &
$\lesssim C_hN_sN_\phi^2 + C_UN_sN_\phi^4$ \\
\hline
\end{tabular}
\end{adjustbox}
\caption{Summary of the resource requirements for encoding the $\phi^4$ Hamiltonian in the harmonic-oscillator basis in momentum space ($H_p$) and coordinate space ($H_x$), using binary and unary boson-to-qubit encodings. The coordinate-space representation reduces the number of Pauli strings from quartic to linear scaling with the number of sites (or modes), up to the effective bandwidths $C_h$ and $C_U$.}
\label{tab:sumup}
\end{table}
The results presented in this paper demonstrate that the choice of basis provides an additional lever for reducing the cost of quantum simulations. While boson-to-qubit mappings determine how the local Hilbert space is encoded through the implementation of the creation and annihilation operators, the choice of representation influences the structural properties of the Hamiltonian itself, such as its sparsity and locality. These two aspects are largely independent and should therefore be viewed as complementary optimization strategies. In the present case, changing from the harmonic-oscillator $\mathrm{HO}_p$ to the $\mathrm{HO}_x$ representation reduces the scaling of the number of Pauli strings from quartic to linear in the number of lattice sites (up to the effective bandwidths). This reduction is a consequence of the change of basis and is accompanied by a smaller Pauli $1$-norm over a broad range of parameters.

The comparison between the $\mathrm{HO}_p$ and $\mathrm{HO}_x$ representations also highlights a trade-off between the free and interacting parts of the Hamiltonian. In the $\mathrm{HO}_p$ formulation, the quadratic Hamiltonian is diagonal whereas the interaction is delocalized over all momentum modes satisfying momentum conservation. In contrast, the $\mathrm{HO}_x$ representation yields an approximately band-diagonal free Hamiltonian together with a localized interaction. Consequently, the benefit of the $\mathrm{HO}_x$ representation depends on the physical regime. For weakly interacting theories, where the free Hamiltonian dominates, the diagonal structure of $\mathrm{HO}_p$ remains advantageous. As the interaction strength increases, the locality of the interaction in the $\mathrm{HO}_x$ representation outweighs the loss of an exactly diagonal free Hamiltonian, leading to lower implementation costs. Moreover, increasing the bare mass further enhances this locality by reducing the effective bandwidths of both the hopping matrix and the interaction tensor.

These structural properties have direct consequences for quantum algorithms. A smaller number of Pauli strings reduces the measurement overhead of algorithms based on Pauli decomposition, while a smaller Pauli $1$-norm improves the query complexity of block-encoding and qubitization-based Hamiltonian simulation algorithms. Since these improvements originate from the Hamiltonian representation rather than from a particular algorithm, they are expected to benefit a broad class of quantum simulation methods.

\section{Conclusion and perspectives}

\label{sec6}
In this paper, we have thus introduced and analyzed in detail the harmonic-oscillator basis formulation of the lattice $\phi^4$ Hamiltonian in coordinate space. By studying the structure of the resulting one-body matrix and interaction tensor, we have shown that the Hamiltonian exhibits an effective locality in this representation. This property allows controlled bandwidth truncations while preserving the low-energy spectrum, as confirmed by comparison with the standard harmonic-oscillator momentum-space formulation.

Furthermore, we have investigated the implications of this locality for quantum simulation by estimating the resources required after unary and binary boson-to-qubit encodings. Compared with the momentum-space representation, the coordinate-space formulation reduces the number of Pauli terms and leads to a smaller Pauli $1$-norm over a range of parameters. These results show that the choice of basis can significantly influence the cost of quantum simulations of field theories.

Several directions naturally follow from the present paper. A first objective is to assess the practical impact of the harmonic-oscillator coordinate-space representation on quantum algorithms through explicit simulations. In particular, it would be interesting to compare the performance of quantum eigensolvers in the momentum- and coordinate-space formulations. Promising candidates include quantum Krylov and quantum Lanczos methods, as well as the broader class of quantum subspace algorithms (see for example \cite{MRKrylov, QSM, QKrylov, QLanczos}), which have recently been applied to lattice gauge theories \cite{KrylovLGT}.

A second direction is the investigation of alternative bases. While the harmonic-oscillator basis provides a natural representation both in momentum and coordinate space, other localized representations may lead to even sparser Hamiltonians. In particular, wavelet bases, which combine locality with a multiscale description of the fields, constitute promising candidates for quantum simulations of quantum field theories.

\bibliographystyle{unsrt} 
\bibliography{Lanczos-QFT.bib}
\end{document}